\documentclass[twoside,12pt]{article}
\usepackage{epsfig}

\newcommand{\be}{\begin{equation}}
\newcommand{\mat}[1]{ \vec{\vec{#1 }}_{i j}}
\newcommand{\ave}[1]{\left\langle #1 \right\rangle}
\newcommand{\ee}{\end{equation}}
\newcommand{\bea}{\begin{eqnarray}}
\newcommand{\eea}{\end{eqnarray}}

\begin{document}

\title{ \vspace{1cm}Bulk viscosity-driven freeze-out in heavy ion collision }
\author{
Giorgio Torrieri$^{1,2}$,Igor Mishustin$^{2,3}$,Boris Tom\'a\v{s}ik$^{4,5}$\\
$^1$Institut f\"ur Theoretische Physik, J.W. Goethe-Universit\"at, Frankfurt, Germany\\
$^2$FIAS,  J.W. Goethe-Universit\"at, Frankfurt, Germany\\
$^3$Kurchatov Institute, Russian Research Center, Moscow 123182, Russia.\\
$^4$Univerzita Mateja Bela, Bansk\'a Bystrica, Slovakia \\
$^5$Czech Technical University in Prague, Prague, Czech Republic
}
\maketitle
\begin{abstract}
We give an review the HBT puzzle, and argue that its resolution requires the introduction of new physics close to the phase transition scale.   We argue that a candidate for this new physics is bulk viscosity, recently postulated to peak, and even diverge, close to the phase transition temperature.  We show that such a viscosity peak can force the system created in heavy ion collisions to become unstable, and filament into fragments whose size is weakly dependent on the global size of the system, thereby triggering freeze-out.
\end{abstract}
\section{The HBT puzzle\label{Int}}
\begin{figure}[t]
\epsfig{file=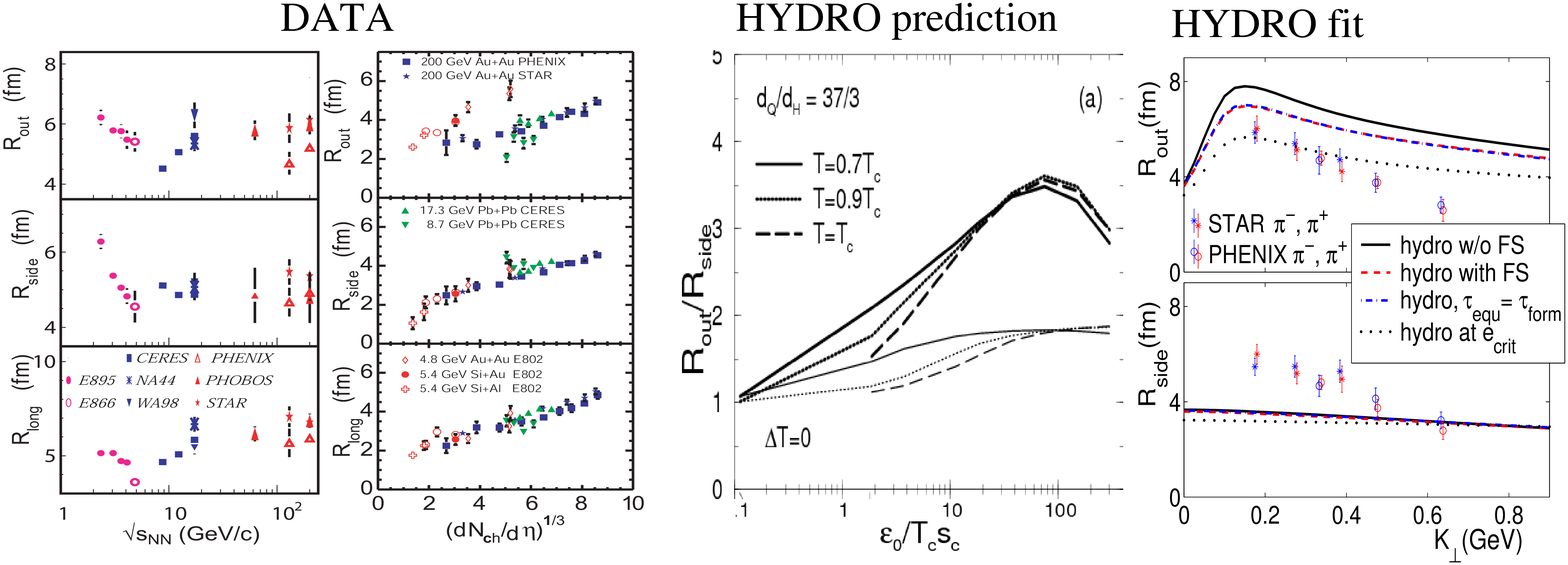,width=17cm}
\caption{ Dependence of  HBT radii on multiplicity, across different energies \cite{hbtreview} (left panel), an expectation from hydrodynamics \cite{predictions} (center panel), and a fit with hydrodynamic model \cite{hydroheinz} (right panel)
\label{hbtstinks} 
}
\end{figure}

One of the most unexpected, and as yet unexplained, experimental results found at the Relativistic Heavy Ion Collider (RHIC) concerns the description of particle interferometry observables \cite{hbtreview}.   It was originally expected that the deconfined matter would be a highly viscous, weakly interacting quark gluon plasma \cite{danielgyul}.   Thus, ideal hydrodynamics would not provide a good description of flow observables sensitive to the early stages of the collision, such as azimuthal anisotropy.   The signature of choice of a phase transition from hydrodynamics models, one less sensitive to viscosity, was to be an increase of the ``out'' to ``side'' emission radius  ratio
($R_o$ and $R_s$, see Fig. \ref{hbtstinks} center panel) due to longer lifetime of the system, caused by the softening of the equation of state in the transition/crossover region \cite{predictions}.   

The data, however, exhibited an opposite behaviour.   Hydrodynamic simulations provided a good description of transverse momentum spectra and their azimuthal anisotropy.   The same simulations, however, failed to describe HBT data \cite{hydroheinz}.
Perhaps the most surpising aspect of the problem is the {\em way} in which the data does not fit:
Measured parameters $R_{o}$ and $R_{s}$ are  nearly identical.  (Fig. \ref{hbtstinks} left and right panel)
Their (positive) difference $R_{o}^2 - R_{s}^2$ is thought to 
depend on the duration of particle emission.
Hence, it looks like the fireball emits particles almost instantaneously and does not 
show any sign of phase transition or crossover. Hydrodynamics, with ``reasonable'' freeze-out condition (such as a freeze-out
temperature of 100 MeV or so) can not describe this even qualitatively.   This behaviour, when compared to lower energy data, exhibits remarkably good scaling with multiplicity (Fig. \ref{hbtstinks} left panel).   The scaling's {\em existance}, however, is by itself surprising since the QCD equation of state, with it's critical density for a phase transition, should break it.

There could be three possible approaches to the HBT puzzle.  It could be that the system is simply too complicated, 
and that once we include all possible improvements (full 3D calculation, viscosity, hadronic kinetic afterburner, in-medium hadron modifications,pre-existing flow etc.), everything will fit.
It could also be that we are drastically misunderstanding the data, and the HBT puzzle is a 
symptom of inapplicability of hydrodynamics to heavy ion collisions.
Finally, it could be that the hydrodynamic approach is basically correct, but just one element 
of relevant physics is omitted  \cite{ourbulk}.   

The second possibility is unlikely because, in some ways, hydrodynamic prescription {\em does} fit HBT data.   
Scaling of HBT radii with the multiplicity rapidity density $(dN/dy)^{1/3}$, over a large range 
of energies \cite{hbtreview} is typical for an isentropically expanding fluid that suddenly breaks apart.
In addition, the good description, within parameters compatible with what is needed to describe flow, of the {\em azimuthal} dependence of HBT radii \cite{hydroheinz}, also suggests that the hydrodynamic framework is a good ansatz for describing the matter produced in heavy ion collisions {\em up to freeze-out}.

The first possibility also appears problematic:
successful models and/or parametrisations of the freeze-out which describe HBT radii are found in the literature 
\cite{models}, and they could provide a way to gain insights into what is missing. However, we feel that  successful description involves a dynamical description from initial conditions {\em plus a freeze-out criterion}, rather than a fit to data with assumptions put in ``by hand''.   Such a description is so far lacking.
Furthermore, the most plausible refinements to hydrodynamics, namely
implementation of fully three-dimensional models and
the addition of a kinetic theory afterburner \cite{hirano} do not do anything
to solve the HBT discrepancy, suggesting that the problem is not refinements but rather one large missed physical effect

\section{Bulk viscosity close to $T_c$}
The bulk viscosity of high temperature strongly interacting matter has recently been calculated using perturbative QCD \cite{amybulk}, and found to be negligible, both in comparison to shear
viscosity and w.r.t. its effect on any reasonable collective evolution of the system.
This is not surprising:  The QCD Lagrangian, as long as no ``heavy'' quarks are present, is nearly conformally invariant \cite{amybulk}.    Since, within a
fluid, the violation of conformal symmetry is linearly proportional to
a bulk viscosity term, the near conformal invariance of the QCD
Lagrangian should guarantee that bulk viscosity is nearly zero,
\textit{in the perturbative regime}.   In the hadron gas phase, the numerous scales associated
with hadrons render conformal invariance a bad symmetry, and hence it
is natural  that bulk viscosity not be negligible. 

This is, again, rooted in a fundamental feature of QCD:  the
non-perturbative \textit{conformal anomaly}, that manifests itself in the
scale (usually called $\Lambda_{QCD}$) at which the QCD coupling
constant stops being small enough for the perturbative expansion to
make sense.  This scale coincides with the scale at which confining forces hold hadrons together.
This violation of conformal invariance is not seen perturbatively, but
should dominate over the perturbatively calculated bulk viscosity as temperature drops
close enough to the QCD phase transition.
What happens to \textit{bulk} viscosity in this regime, where hadrons are not yet
formed, presumably the matter is still deconfined, but conformal
symmetry is badly broken?  There is a quenched lattice calculation, and compelling physical arguments \cite{bulkvisc} 
that bulk viscosity rises sharply, or even diverges, close to the phase transition temperature.

Lattice simulations find that $T^{\mu}_{\mu}$ (=0 for a
conformally invariant system), increases  rapidly close to
$T_c$.  Remembering that the shear ($\eta$) and bulk ($\zeta$) viscosities
roughly scale as \cite{weinberg}
\begin{eqnarray}
\label{bulkgeneral}
\eta  \sim  \tau_{\rm elastic} T^4 \phantom{AAAA}
\zeta  \sim  \left( \frac{1}{3} - v_s^2 \right)^2 \tau_{\rm inelastic} T^4
\end{eqnarray}
where $\tau_{\rm (ine)elastic}$ refers to the equilibration timescale of (ine)elastic
collisions.   Finite temperature sum rules in conjunction with lattice data \cite{bulkvisc} give a sharp rise in bulk viscosity.

The rise is, in fact, likely to be considerably sharper than \cite{bulkvisc} suggests.    The dependence of $\tau_{\rm inelastic}$ on temperature can be guessed from
the fact that, at $T_c$, the quark condensate $\ave{q \overline{q}}$ 
acquires a finite value, and the gluon condensate
$\ave{G_{\mu \nu} G^{\mu \nu}}$ sharply increases at the phase
transition.   ``Kinetically'', therefore, timescales of processes
that create extra $q \overline{q}$ and $G G$ pairs should diverge
close to the phase transition temperature, by analogy with the divergence of the spin correlation length in the Ising model close to the phase transition. 
The sharp rise of bulk viscosity can also be understood within string kinetics:
confinement, microscopically, can be thought of
as a ``string tension'' appearing in the potential.      Even in a regime where
the momentum exchange of the average collision is more than enough to
break the string, and the relevant degrees of freedom are still quarks, not mesons, string tension introduces a huge change in kinetics: what were elastic collision without string tension, become inelastic onece string tension appears.
Even if the energy needed to break the string is low, over many collisions, the heat energy would
be converted into creating more slightly colder, less pressing particles.
That's exactly the kind of processes that contribute most to bulk viscosity \cite{jeonvisc}.

These arguments give evidence to the conjecture that, close (from
above) to $T_c$, bulk viscosity goes rapidly from a
negligible value to a value capable of \textit{dominating} the
collective
evolution of the system.
For our analysis, we assume the bulk viscosity to be of the form
$\zeta = s \left(  z_{pQCD} + \frac{z_0}{\sqrt{2\pi}\sigma} \exp(- \frac{t^2}{2 \sigma^2})\right)$
where $t=T-T_c$ and $\sigma=0.01 T_c$ and $z_{pQCD} \sim 10^{-3}$ \cite{amybulk}.  
At $T>T_c$, this ansatz provides a reasonable fit to the results of 
\cite{bulkvisc}, considering the peak height and width of the distribution are still unknown.

\section{Bulk viscosity-triggered fragmenting}
To study the effect of our conjectured behaviour of bulk viscosity on solutions to hydrodynamics, we perform a stability analysis \cite{stability} of a boost-invariant solution to the Navier-Stokes equations (both the 1D and 3D cases).  
The Navier-Stokes equations with boost-invariant symmetry \cite{danielgyul,stability,bjorken} can be rewritten in terms of the Reynolds number $R$, the entropy $s$ the co-moving time $\tau$, the total number ($N$) of dimensions, and the dimensionality of the homogeneous expansion ($M$)
\begin{equation}
\tau^{-M} \frac{ d (\tau^M s)}{d\tau} = \frac{M s}{R\tau}
\end{equation}
$M=1$ $N=3$ corresponds to Bjorken hydrodynamics \cite{bjorken}, $M=N=3$ to a homogeneus 3D expansion.
The Reynolds number is a function of temperature $T$, bulk and shear viscosity $\zeta$ and $\eta$ and entropy density 
$s$:
$R^{-1}= (2(1-M/N)\eta+M\zeta)/(T s \tau)$.
With given expressions for $s(T),\, \eta(T),\, \zeta(T)$, this set of equations becomes closed and solvable.  For the equation of state, we use a parameterization of lattice data (We have checked that our results do not vary qualitatively if the ideal EoS is used).

We follow the stability analysis performed in \cite{stability}.  The amplitude of a generic perturbation to the 1D Boost-invariant system is a vector $\vec{x}$ in the two dimensional space of entropy perturbations and flow (rapidity $y$) perturbations, and its frequency in rapidity can be decomposed into Fourier components
\begin{eqnarray}
\vec{x}(y) = \sum_k \vec{x}(k) e^{ i k y}\phantom{AA},\phantom{AA} x_1=\frac{\delta s}{s} \phantom{AA},\phantom{AA} x_2 = y-y_{spacetime}
\end{eqnarray}
The equation of motion for $\vec{x}$ will then be given by
\begin{equation}
\label{perturbeq}
\tau \frac{\partial}{\partial \tau}  \left(  \begin{array}{c} x_1 \\i x_2  \end{array} \right)
 = \left( \begin{array}{cc}
A_{11} & A_{12}\\
A_{21} & A_{22}
\end{array} \right)
 \left(  \begin{array}{c} x_1 \\i x_2  \end{array} \right)
\end{equation}
where $\mat{ A}$ is a real matrix function of $s,R$ and $k$.  We refer the reader to Eq. 4.23-4.26 of \cite{stability} for the full form of $\mat{ A}$.  The system's stability can be understood via the behavior of the modulus of $\vec{x}$, $X = \vec{x}^T \vec{x}$
\begin{eqnarray}
\tau \frac{\partial}{\partial \tau} X = \vec{x}^T \mat{ M}  \vec{x}\phantom{AA},\phantom{AA} \mat{M}=\mat{A}+\mat{A}^T 
\end{eqnarray}
Since $\mat{ M}$ is real and symmetric, it will always have two real eigenvalues, $\lambda_{max}$ and $\lambda_{min}$ (corresponding eigenvectors $\vec{x}_{max,min}$), as well as orthogonal matrices $\mat{B} $ diagonalizing it.    Defining $\vec{y}_i = \mat{B}^{-1} \vec{x}_j$we see that
$\lambda_{min} y_{min}^2  <\tau \frac{\partial X}{\partial \tau} < \lambda_{max} y_{max}^2$
Thus, if $\lambda_{min}>0$, the system is unstable, since perturbation in any direction will produce a positive growth rate.  An instability will grow as a power-law with $\tau/\tau_{ini}$, where $\tau_{ini}$ is the starting time of the perturbation
$\vec{x}(\tau) \sim  ( {\tau}/{\tau_{ini}} )^{\lambda} \vec{x}(\tau_{ini})$.  If 
 $\lambda_{max}<0$, on the other hand, the system will be stable against  perturbations, an instability will be suppressed as $\sim ( {\tau}/{\tau_{ini}} )^{\lambda(<0)}$.

If $\lambda_{min}<0$ and $\lambda_{max}>0$ some modes will be stable and some will be unstable.   
In the latter case, the time dependence of $\mat{A}$ will in general continuously rotate $\vec{x}_{min}$ and  $\vec{x}_{max}$ in time, stopping the growth of the instability: Since $\mat{A}$ is time-dependant, $\vec{x}_{max}(\tau_{ini})$ might be in the direction of  $\vec{x}_{min}(\tau>\tau_{ini})$ a short time later.
Solving Eq. \ref{perturbeq} will take this effect into account.

In what follows, we use $z_0=0,0.1,1$  $T_c$ 
 and evolve the system from an initial temperature $T=0.3$ 
GeV and comoving time 0.6 $fm$. Note that the qualitative features of our study are independent of the details of the evolution before $T_c$, such as 
the initial temperature and timescale. 

\begin{figure}[h]
\epsfig{width=17cm,clip=1,figure=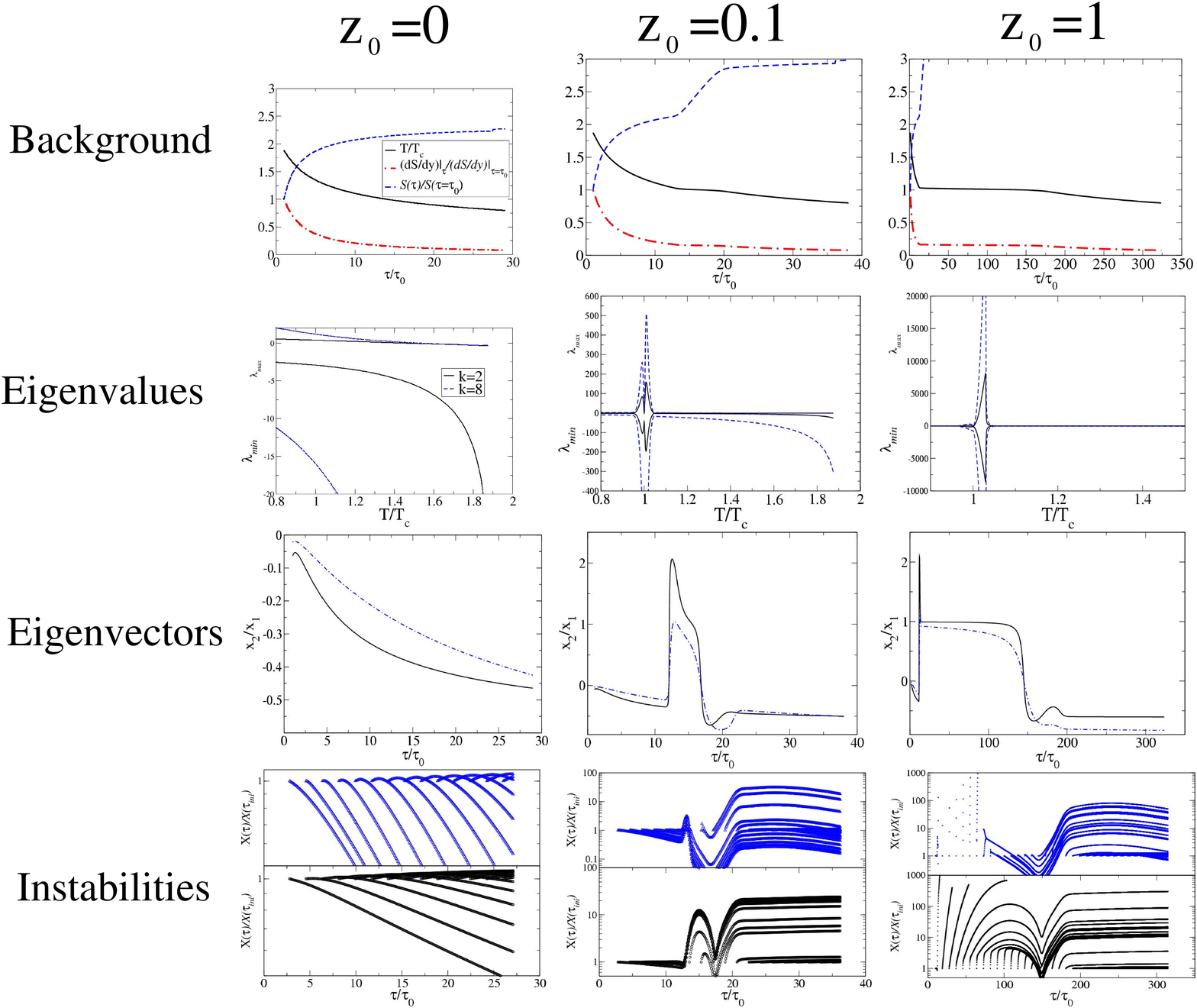}
\caption{\label{allplots}  Instability analysis of the boost-invariant solution. Top panel shows the evolution of the background temperature (solid), entropy density (dot-dashed) and total entropy (dashed).  The two middle panels show the eigenvalues and eigenvectors of the $k=2$ (solid) and $k=8$ unstable mode.  The bottom panel shows the full evolution of the linearized instability equation of motion.  The three columns represent various appearances of the viscosity peak.  See text for details.}
\end{figure}
Figure \ref{allplots} (upper panel) shows the temperature, entropy density and total entropy in the central rapidity unit as a function of time.   As soon as $z_0$ becomes non-negligible (i.e. viscous forces dominate around $T_c$), the kinematic evolution of the system ``freezes''.
The system then stays at nearly constant temperature, through it keeps producing more and more entropy at the expense of advective energy.   
At first sight, large values of $z_0$ are excluded by HBT data and multiplicity measurements.   However, we will show that this long phase is unstable against small perturbations.   Thus, its further evolution will not be given by the background solution, but by a rapid formation of local inhomogeneities.

Figure \ref{allplots} (second panel) shows the $\lambda_{min}$ and $\lambda_{max}$ Eigenvalues corresponding to representative $k=2,8$ (other values of $k$ were checked not to vary significantly wrt those presented here).   As can be seen, the peak in bulk viscosity forces the growing/damping rates to increase rapidly.   Thus, any initial perturbation in the unstable direction will rapidly grow to a value comparable with the background, unless the system's evolution will stop the growth by rotating the direction of the unstable modes.   

Figure \ref{allplots} (third panel) examines whether this occurs for larger values of $z_0$.   If the peak of viscosity is negligible, the unstable eigenvector keeps rotating throughout the evolution of the system.   Thus, even an unstable mode's growth will very quickly stop growing since the dynamics will turn it into a damped mode.   
When $z_0$ dominates, however, when $T$ approaches $T_c$, the direction of the unstable modes experiences an abrupt rotation.     Then it stays constant throughout the time the system travels through the viscosity peak (this time increases strongly as $z_0$ increases), and gets rotated again as the peak is passed.  The reason for this behavior is clear from the ``freeze'' of the background evolution in the top panel.  Thus, at large $z_0$ unstable modes have all the required time to grow.

Finally, Fig. \ref{allplots} (bottom panel) shows the explicit solution of Eq. 
\ref{perturbeq}.   At each 
time-step, a perturbation is born in the unstable eigenvector mode, and then evolved until the end of the evolution of the system. The plot shows $Z=X(\tau)/X(\tau_{ini})$,  the ratio of strength of the perturbation to the initial strength as a function of time (Note that the starting value is always unity). A large $Z$ does not mean the evolution equation is invalid: For any point in the graph, there will be a small enough perturbation amplitude that survives as a perturbation in the subsequent evolution.  The probability of a {\em larger} perturbation forming and significantly modifying the background, however, should grow strongly with $z_0$.   
It is clear that any microscopic mechanism seeding instabilities at the scale $Z \sim 10^{-1}$,uniformly distributed in $\vec{x}$ and at a rate of  $\sim fm^{-4}$ is likely to generate power-law growing instabilities  a few $fm$ after $T \sim T_c$.  These instabilities should reach $Z \ge 1$, and hence play an important role in the subsequent evolution of the system, a few $fm$ after that. 

The resulting scenario is phenomenologically similar to the fragmenting induced by supercooling in a first order transition \cite{firstorder}, but originating from hydrodynamic rather than thermodyamic instabilities.

We do not expect second order hydrodynamics \cite{israelstewart} to play a big role in {\em starting} the instabilities:
As argued in \cite{ourbulk}, as long as the system's expansion rate as the system approaches $T_c$ is smaller than the relaxation time $\tau_{\Pi}$,
$\tau_\Pi \frac{1}{\sigma}\frac{dT}{d\tau} < 1$
second order terms will not prevent viscous corrections to the pressure of the order of $\zeta/\tau$, but merely localize their propagation.
The (admittedly unreliable) estimates from strongly coupled CFT \cite{janik} suggest that this criterion is amply satisfied, especially considering that the
increase in viscosity causes the background solution to slow down over a timescale much bigger than $\tau_\Pi$.
Second order hydrodynamics might however play a dominant role in stopping the instabilities once they grow to a value comparable to the background.

Recent numerical simulations, with Israel-Stewart hydrodynamics, provide 
evidence that this description is true, as cluster-like instabilities 
seem to be present (Fig. 11 of \cite{denicol})

\section{Fragmenting and the HBT puzzle \label{hbt_fragments}}
HBT radii are directly related to
the system's spacetime correlation tensor  \cite{hydroheinz,hbtreview}\footnote{Here $l$ (``long'') is the z direction (parallel to the
beam), $o$ (``out'') is the direction of the pair momentum, and $s$ (``side'') is the
cross product of the previous two.  Averaging is done over the emission function $S(x,p)$, the probability of producing a particle of momentum $p$ at $x$.  See \cite{hbtreview} for details}
\begin{eqnarray}
R_o^2(K) &=& \ave{(\Delta x_o)^2} - 2 \frac{k_T}{k_0} \ave{\Delta x_o
  \Delta t} + \left(  \frac{k_T}{k_0} \right)^2  \ave{(\Delta t)^2} \label{rout}
\end{eqnarray}
while $R_s^2(K)$ is simply given by $ \ave{(\Delta x_s)^2}$.
Here the $K$ is the sum of the two four-momenta of the pion pair.
As remarked in \cite{hydroheinz}, the $R_o
\sim R_s$ result is not easy to reconcile with naive hydrodynamics
plus a straight-forward (critical temperature) emission for several reasons;

Firstly, the higher the energy, the longer the emission  time,
the larger is the expected discrepancy between $R_o$ and $R_s$.  If the
system starts close to the mixed phase, the timescale of freezing out
should be longer still due to the softest point in the equation of state.
Hence, a generic prediction from Eq. (\ref{rout}) is that 
$R_o/R_s>1$, largely increases with energy, and exhibits a peak when the energy 
density is such that the system starts within, or slightly above the mixed phase.  This is in direct contrast with
experimental data, where $R_o/R_s \simeq 1$ is a feature at all reaction
energies.

Secondly, generally 
$\ave{\Delta x \Delta t}<0$, since particles on
the outer layer freeze-out first.   This increases
$R_o/R_s$ further (cf.\ eq~\ref{rout}).  
Time dilation due to transverse flow does reverse this dependence \cite{hydroheinz}.

Fragmentation of the bulk could help solving 
these problems.  Firstly, fragment size, density and decay timescale, is approximately independent of either
reaction energy or centrality. 
Hence, the near
energy independence of the (comparatively short) emission timescale,
and hence of $R_{o}/R_{s}$, should be recovered.
Secondly, if the decay products do not interact (or do not interact much)
after fragment decay, it can also be seen that $\ave{\Delta x \Delta t}$
can indeed be positive:  outward fragments are moving faster,
resulting in time dilation.   
This effect can be offset by time dilation of
fragment decay by increasing the temperature at which fragments form,
or by increasing fragment size.
Recovering the  linear scaling of the radii with $(dN/dy)^{1/3} (\sim
N_{\rm fragments})$  \cite{hbtreview}, 
while maintaining the correct $R_{o}/R_{s}$ is also possible if
the fragments decay when their distance w.r.t.\ each other is
still comparable
to their intrinsic size.

The bulk-viscosity-driven freeze-out adds another parameter to ab initio HBT calculations:  
in addition to critical temperature/energy density, we now have the fragment size.
To see whether this helps solving the HBT problem, output from hydrodynamics with a high ($T \sim T_c$) freeze-out temperature should be fragmented into piecess with a certain distribution in size, 
which then produce hadrons according to the prescription in the Appendix of \cite{ourbulk}.

In conclusion, we have introduced the HBT puzzle, and explained how bulk viscosity could help solve it by triggering the fragmenting of the system close to the critical temperature.  We hope that, in the near future, this scenario will be developed to the point of experimental falsification.

GT would like to thank the Alexander von Humboldt Foundation and the 
LOEWE foundation for 
the support provided, and to Mike Lisa, Sangyong Jeon, Guy Moore and Johann Rafelski for fruitful discussions.
BT acknowledges support from
VEGA 1/4012/07 (Slovakia) as well as MSM 6840770039 and LC 07048
(Czech Republic).
IM acknowledges support provided by the DFG grant 436RUS 113/711/0-2
(Germany) and grants RFFR-05-02-04013 and NS-8756.2006.2 (Russia).
GT is grateful to the Erice school and its organizers for the fellowship that allowed him to attend the school.


\end{document}